\documentclass[aip,reprint]{revtex4-2}

\usepackage{color}
\usepackage{graphicx}
\usepackage{amsmath}
\usepackage{subcaption}

\usepackage{natbib}

\begin{document}
\title{Magnetized ICF implosions: Non-axial magnetic field topologies}

\author{C. A. Walsh}
\email{walsh34@llnl.gov}
\affiliation{Lawrence Livermore National Laboratory, 7000 East Avenue, Livermore, CA 94550, USA}
\author{D. J. Strozzi}
\affiliation{Lawrence Livermore National Laboratory, 7000 East Avenue, Livermore, CA 94550, USA}
\author{A. Povilus}
\affiliation{Lawrence Livermore National Laboratory, 7000 East Avenue, Livermore, CA 94550, USA}
\author{S. T. O'Neill}
\affiliation{Imperial College, Exhibition Rd, South Kensington, London SW7 2AZ, UK}
\author{L. Leal}
\affiliation{Lawrence Livermore National Laboratory, 7000 East Avenue, Livermore, CA 94550, USA}
\author{B. Pollock}
\affiliation{Lawrence Livermore National Laboratory, 7000 East Avenue, Livermore, CA 94550, USA}
\author{H. Sio}
\affiliation{Lawrence Livermore National Laboratory, 7000 East Avenue, Livermore, CA 94550, USA}
\author{B. Hammel}
\affiliation{Lawrence Livermore National Laboratory, 7000 East Avenue, Livermore, CA 94550, USA}
\author{B. Z. Djordjevi\'c}
\affiliation{Lawrence Livermore National Laboratory, 7000 East Avenue, Livermore, CA 94550, USA}
\author{J. P. Chittenden}
\affiliation{Imperial College, Exhibition Rd, South Kensington, London SW7 2AZ, UK}
\author{J. D. Moody}
\affiliation{Lawrence Livermore National Laboratory, 7000 East Avenue, Livermore, CA 94550, USA}

\date{\today}

\begin{abstract}
	This paper explores 4 different magnetic field topologies for application to spherical inertial confinement fusion implosions: axial, mirror, cusp and closed field lines. A mirror field is found to enhance the impact of magnetization over an axial field; this is because the mirror field more closely follows the hot-spot surface. A cusp field, while simple to generate, is not found to have any benefits over the tried-and-tested axial field. Closed field lines are found to be of the greatest benefit to hot-spot performance, with the simulated design undergoing a 2x increase in ion temperature before $\alpha$-heating is considered. The plasma properties of the simulation with closed field lines are radically different from the unmagnetized counterpart, with electron temperatures in excess of 100 keV, suggesting that a fundamental redesign of the capsule implosion is possible if this method is pursued.
\end{abstract}
\maketitle

\begin{figure*}
	\centering
	\centering
	\includegraphics[scale=0.9]{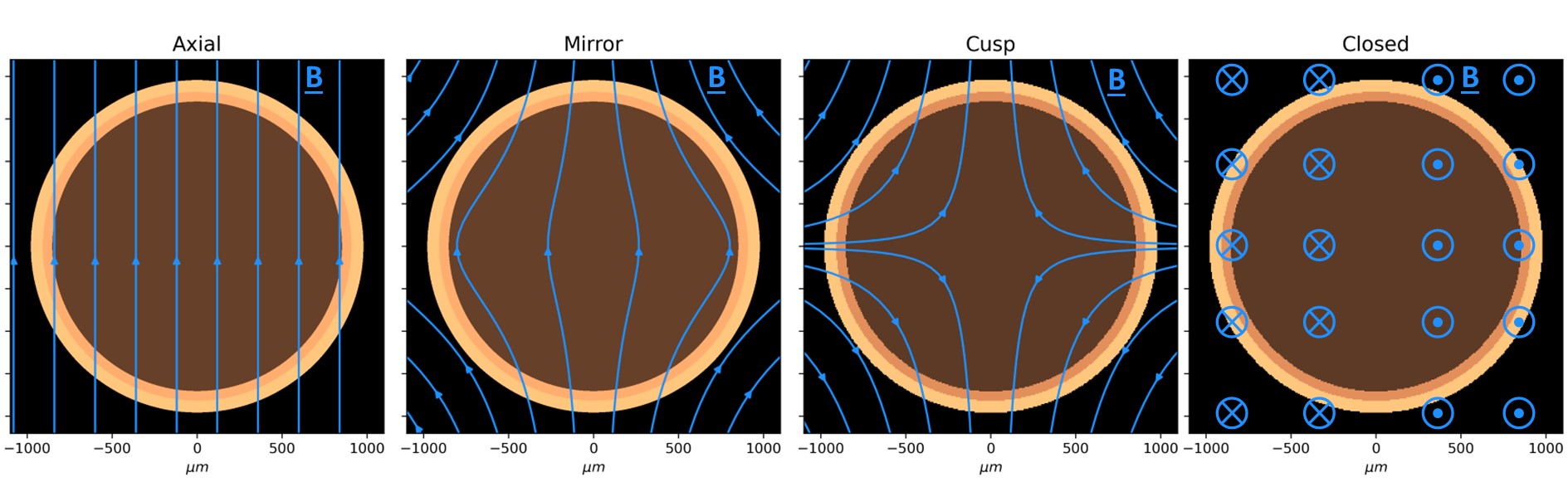}\caption{\label{fig:B0} Initial magnetic field streamlines plotted over the capsule density for the 4 topologies in this paper.}	 
\end{figure*}

Axial magnetic fields have been utilized on spherical inertial confinement fusion (ICF) implosions, with increased performance observed on both the OMEGA laser facility\cite{changFusionYieldEnhancement2011} and the National Ignition Facility (NIF)\cite{moodyIncreasedIonTemperature2022}. The enhancement in hot-spot temperature is expected to be predominantly from a reduction in heat conducted away from the fusing fuel \cite{perkinsPotentialImposedMagnetic2017,walshMagnetizedICFImplosions2022}, although magnetization of the $\alpha$-particles will become more significant at higher yields. Magnetic fields have also been shown to directly suppress perturbation growth through magnetic tension \cite{srinivasanMitigatingEffectMagnetic2013,walsh2021magnetized,walsh2024}.

However, there are a number of signs that axial magnetic fields (which are cylindrically symmetric) are not the optimum topology for a spherical implosion. First and foremost, an axial field affects the capsule poles different to the waist \cite{walshPerturbationModificationsPremagnetisation2019}, resulting in an inherently asymmetric implosion; the heat from the stagnated hot-spot propagates unrestricted towards the poles. For the low yield implosions fielded to date, it is thought that this effect can be counteracted by driving the implosion asymmetrically \cite{walshMagnetizedICFImplosions2022}. It may prove more difficult to account for this inherent asymmetry in the high yield regime that has recently been achieved on the NIF \cite{abu-shwarebLawsonCriterionIgnition2022,oneill2024burnpropagationmagnetizedhighyield}. For direct-drive, there are additional issues where an axial magnetic field is predicted to enhance laser drive asymmetries \cite{walshMagnetizedDirectlydrivenICF2020}, with experiments providing some evidence of this effect \cite{boseEffectStronglyMagnetized2021}. 

In addition, simulations and theory predict that the benefit of magnetization has plateaued at the 26T maximum field strength so far fielded \cite{walshMagnetizedICFImplosions2022,oneill2024burnpropagationmagnetizedhighyield,strozzi2024} (although this doesn't account for the impact of magnetization on mix, which could enhance the plateau for high B-field strength). Experiments using intermediate field strengths on warm gas targets so far suggest that raising the magnetic field above 26T will have little effect \cite{sio2023}. Perhaps there is another way to increase the impact of magnetization?

This paper investigates alternative magnetic field topologies for use in spherical ICF experiments.

The Gorgon magnetohydrodynamic (MHD) code is utilized for this investigation \cite{ciardiEvolutionMagneticTower2007,chittendenXrayGenerationMechanisms2004,walshSelfGeneratedMagneticFields2017}. The magnetic transport in Gorgon includes bulk plasma advection and Nernst advection down temperature gradients \cite{walshExtendedmagnetohydrodynamicsDensePlasmas2020}. While included in this study, the resistive diffusion of magnetic field and the Biermann Battery generation are found to be of little importance \cite{walsh2024,walshExtendedMagnetohydrodynamicEffects2018}. Updated transport coefficients are used \cite{sadlerSymmetricSetTransport2021,daviesTransportCoefficientsMagneticfield2021}, which have been found to reduce magnetic field twisting in pre-magnetized implosions \cite{2021}. Gorgon simulations compare favorably with magnetic flux compression experiments \cite{10.1088/1361-6587/ac3f25,gotchevLaserDrivenMagneticfluxCompression2009}, improving confidence in the magnetic transport scheme.

The electron heat-flow in Gorgon follows:
\begin{equation}
	\underline{q}_e = -\kappa_{\parallel} \underline{\hat{b}} (\underline{\hat{b}} \cdot \nabla T_e)  -\kappa_{\bot} \underline{\hat{b}} \times  (\nabla T_e \times \underline{\hat{b}}) - \kappa_{\wedge} \underline{\hat{b}} \times \nabla T_e  \label{eq:magheatflow}
\end{equation}
Where $\kappa_{\parallel}$, $\kappa_{\bot}$ and $\kappa_{\wedge}$ are thermal conductivities \cite{braginskiiTransportProcessesPlasma1965,epperleinPlasmaTransportCoefficients1986}. The heat-flow along magnetic field lines ($\kappa_{\parallel}$) is unaffected by the magnetic field. The perpendicular thermal conductivity, on the other hand, is reduced as the plasma magnetization ($\omega_e \tau_e$) increases. The final term is the Righi-Leduc heat-flow, which peaks at approximately $\omega_e \tau_e \sim 0.1-1.0$. As this paper is most concerned with highly magnetized targets, the Righi-Leduc heat-flow is found to be insignificant. The scheme for anisotropic thermal conduction in Gorgon is centered symmetric, which reduces numerical diffusion across magnetic field lines \cite{sharmaPreservingMonotonicityAnisotropic2007}. Recently validated flux limiters applicable to the magnetized regime have been used \cite{walshKineticCorrectionsHeatflow2024}.

Fusion-produced $\alpha$-particles are propagated using a particle-in-cell method, with the magnetic field confining their trajectories \cite{tongAlphaHeatingPowerBalance}. A magnetic field strength of 10 kT restricts a DT-produced $\alpha$-particle to a radius of 27$\mu$m when travelling perpendicular. The impact of the Lorentz force on the bulk plasma is also included in all simulations, but has only a secondary effect on the plasma properties. The Lorentz force becomes more important once the effect of magnetization on perturbation growth is considered \cite{chandrasekharHydrodynamicHydromagneticStability1962,srinivasanMitigatingEffectMagnetic2013,walshMagnetizedAblativeRayleighTaylor2022}; there are no perturbations included in this paper, so there is no impact of the B-field on mix. 

The design chosen to investigate the magnetic field topologies is N170601, a cryogenic DT layered implosion from the HDC campaign \cite{clark2019}. This experiment was the first to exceed a yield of $10^{16}$ neutrons and is chosen due to the $\alpha$-heating being moderate; the self-heating of the hot-spot is mostly turned off to isolate the impact of the magnetic field on the heat-flow.  

Simulations in this paper do not incorporate the hohlraum, using a prescribed frequency-dependent radiation source from a separate calculation. Therefore, these simulations do not include any impact of the magnetic field on the radiation drive. Simulations of hohlraums with an applied magnetic field can be found here \cite{strozziImposedMagneticField2015,strozzi2024,montgomeryUseExternalMagnetic2015}. The radiation drive is taken as symmetric throughout this paper to isolate the impact of magnetic topology on shape.

The 2D Eulerian simulations are first run in spherical geometry for the drive phase, before being restarted into cylindrical geometry for hot-spot stagnation. During the drive phase a radial resolution of 1$\mu$m is used, as well as 360 polar zones. For the stagnation the resolution is 1$\mu$m in both the axial and radial directions.

The magnetic field topologies considered in this paper are axial (section \ref{sec:axial}), mirror (section \ref{sec:mirror}), cusp (section \ref{sec:cusp}) and closed field lines (section \ref{sec:closed}). These topologies are plotted in figure \ref{fig:B0} overlaying the spherical capsule before the radiation drive has begun. The possibility of twisted magnetic field lines (essentially a combination of axial and closed topologies) is briefly discussed in section \ref{sec:twist}.

\section{Axial Magnetic Field \label{sec:axial}} 

\begin{figure}
	\centering
	\centering
	\includegraphics[scale=0.6]{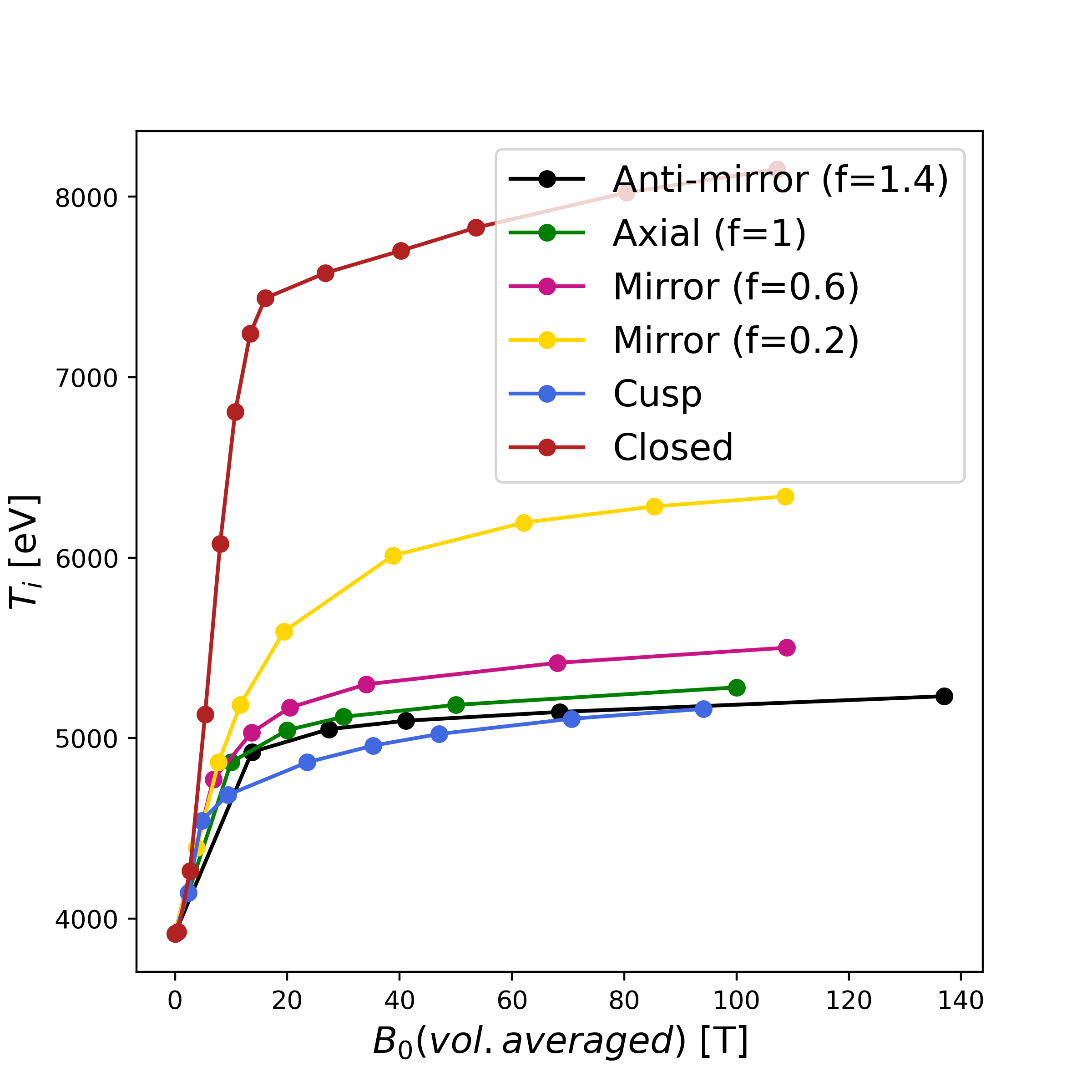}\caption{\label{fig:Tamp_B} Burn-averaged ion temperature against the initial volume-averaged B-field strength in the capsule fuel for various 2D simulations. }	 
\end{figure}

This section revisits the prevailing ideas behind the impact of an initial axial field on thermal conduction in a magnetized implosion. In particular, the scaling of hot-spot performance with applied field strength is sought. $\alpha$-heating is not included in the calculations, unless explicitly mentioned. 

Through simple arguments about ablation of cold fuel into an ICF hot-spot, the following temperature scaling can be formulated \cite{walshMagnetizedICFImplosions2022}:

\begin{equation}
	T \sim \kappa_{\text{eff}}^{-2/7} \label{eq:T}
\end{equation}
Where $\kappa_{\text{eff}}$ is the "effective electron thermal conductivity" and is a dimensionless quantity. This parameter is intended to simplify the inherently 2D behavior of magnetized implosions.

For a purely axial magnetic field applied to a perfectly spherical hot-spot, 1/3 of the hot-spot surface has its normal colinear with the magnetic field direction ($\iint | B_z \cdot \underline{\hat{n}} | dS = 1/3$). This makes intuitive sense, as the magnetic field can only inhibit thermal conduction in 2 of the 3 dimensions and the global heat-flow has no preferred direction. Therefore, the following effective thermal conductivity has been proposed for an axial field \cite{walshMagnetizedICFImplosions2022}:

\begin{equation}
	\kappa_{\text{eff,axial}} = \frac{1}{3} + \frac{2}{3}\frac{\kappa_{\perp}}{\kappa_{\parallel}}\label{eq:kappa_axial}
\end{equation}
Note that this assumes the initial axial field is still axial upon stagnation. 

Figure \ref{fig:Tamp_B} shows the burn averaged ion temperature against applied axial field strength, alongside the other magnetic topologies discussed in this paper. As the field strength is increased, $\kappa_{\text{eff,axial}}$ decreases, and the temperature is enhanced. This effect sees a plateau: for a fully magnetized plasma ($\kappa_{\perp} = 0$), the combination of equations \ref{eq:T} and \ref{eq:kappa_axial} gives a 37\% enhancement in hot-spot temperature. The simulations roughly agree with this, giving a maximum temperature enhancement of 35\%. Note that the scaling assumes the magnetic field has no impact on other energy transport processes, e.g. radiative losses. 

\begin{figure}
	\centering
	\centering
	\includegraphics[scale=0.6]{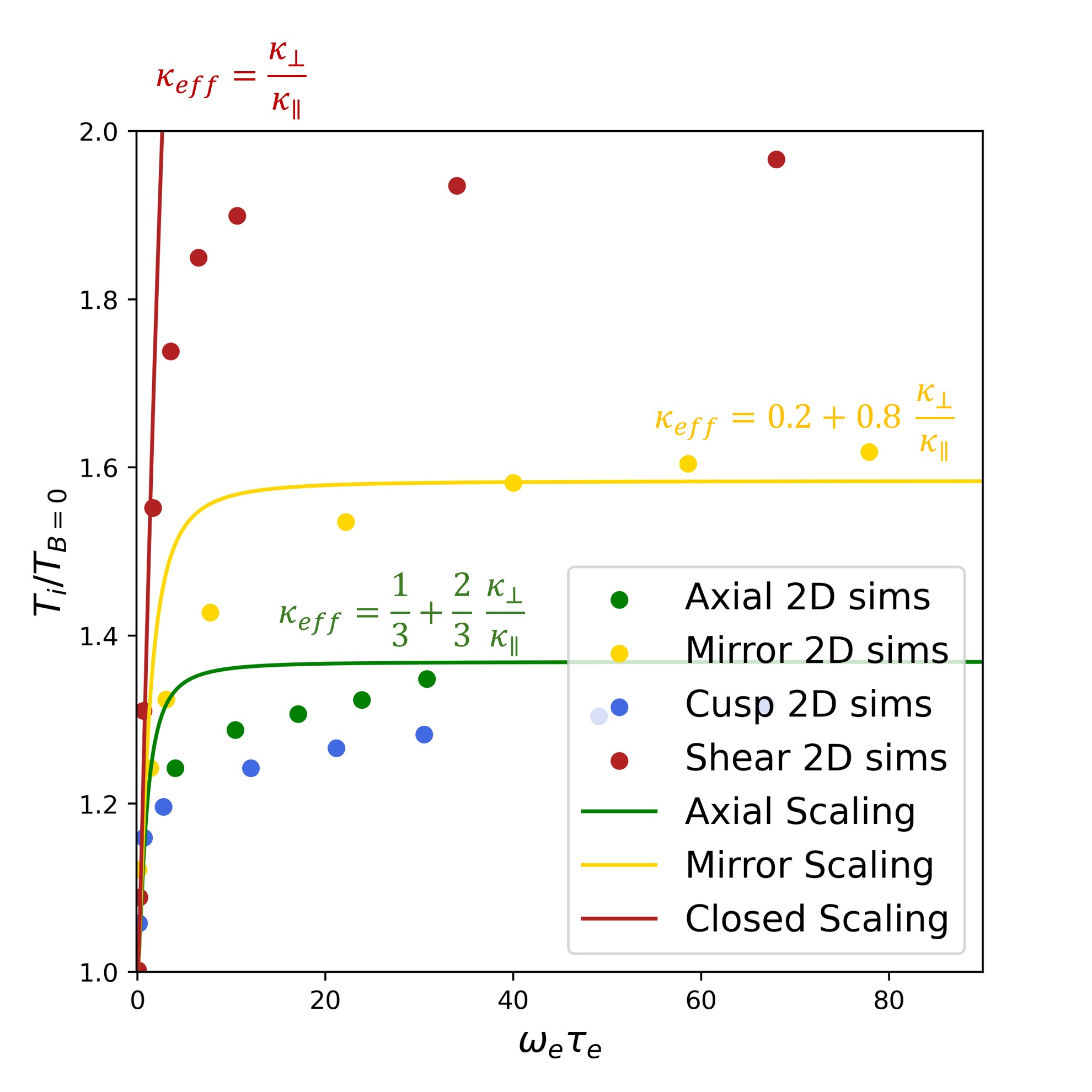}\caption{\label{fig:Tamp_wt} Ion temperature enhancement due to magnetization against the burn-averaged electron magnetization ($\omega_e \tau_e$) for various simulations. Also plotted are several theoretical scalings.}	 
\end{figure}

This result can be represented more universally by plotting the temperature amplification due to a magnetic field ($T_i/T_{i,B=0}$) against the burn averaged electron magnetization in figure \ref{fig:Tamp_wt}. The axial scaling is simply $	\kappa_{\text{eff,axial}}^{-2/7}$ and applies for any spherical ICF implosion design. The theory does a reasonable job of predicting the temperature enhancement for the design chosen in this paper. 

Experiments have corroborated this theoretical prediction. The maximum temperature enhancement from magnetization on the NIF has been 40\% (from a 26T field) \cite{moodyIncreasedIonTemperature2022}, only slightly higher than the 37\% maximum claimed here. There are also signs that the temperature enhancement has plateaued, with a 12T experiment measuring a 30\% temperature enhancement \cite{sioPerformanceScalingApplied2023}, i.e. 3/4 of the effect from less than 1/2 of the applied field strength.

Naturally the temperature enhancement is only of interest to inform how the yield increases. Theory and simulations agree that as the characteristic temperature of the hot-spot increases, the density decreases such that the total hot-spot pressure is roughly unchanged \cite{walshPerturbationModificationsPremagnetisation2019,walshMagnetizedICFImplosions2022}. Once these are taken into account through the fusion rate, the yield enhancement is no longer just a function of $\kappa_{\text{eff,axial}}$, but also depends on the unmagnetized hot-spot temperature, finding that a lower temperature hot-spot has its performance further enhanced by magnetization. Additional details can be found in \cite{walshMagnetizedICFImplosions2022}.

\begin{figure}
	\centering
	\centering
	\includegraphics[scale=0.6]{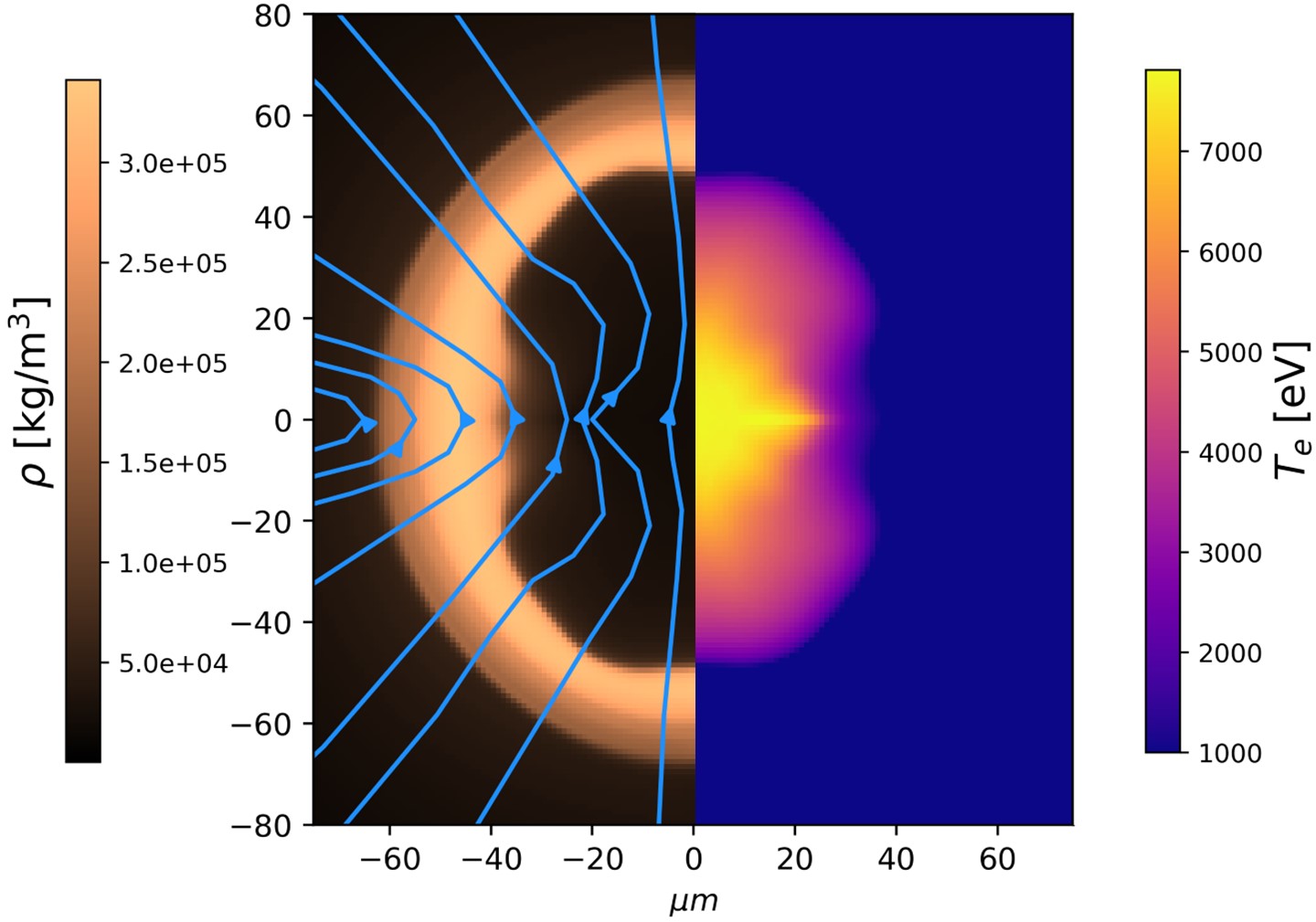}\caption{\label{fig:stag_ax} Density, magnetic streamlines and electron temperature at peak neutron production for a $B_0z = 30T$ axial field.}	 
\end{figure}

An additional complication with understanding the yield enhancement due to a magnetic field is that the yield strongly depends on the stagnated hot-spot shape; this is less important for the temperature enhancement \cite{walshMagnetizedICFImplosions2022}. Figure \ref{fig:stag_ax} shows the stagnated hot-spot density, electron temperature and magnetic field streamlines for a capsule with a 30T initial axial magnetic field (t=8.5ns). The hot-spot has elongated along the magnetic field lines due to enhanced thermal conduction in this direction. It has been found that the theoretical scaling of yield with magnetization only applies once the capsule shape has been corrected to be round \cite{walshMagnetizedICFImplosions2022}. For these reasons, the main figure of merit for the different magnetic field topologies is taken to be the temperature enhancement for the sake of this paper.

\section{Mirror Magnetic Field \label{sec:mirror}}

The advantage of a mirror field is apparent in figure \ref{fig:B0}: the field lines more closely follow the shape of the spherical surface when compared with the axial case. This is better suited to restricting heat-flow out of the hot-spot upon stagnation.

To investigate this, the following form for an initial magnetic field strength is used:

\begin{equation}
	B_z = B_{M0z} \Big(1- (1-f) cos \big( \frac{\pi z}{2 L_z} \big) \Big)
\end{equation}

\begin{equation}
	B_r = B_{M0z} \frac{\pi r_{cyl}}{4 L_z} (1-f) sin \big(\frac{\pi z}{2L_z}\big)
\end{equation}
Where $r_{cyl}$ is radius in cylindrical co-ordinates and $f$ is the parameter that sets the mirror strength. $f=1$ gives an axial field. $f=0.2$ gives a magnetic field strength at the capsule center that is 20\% of the value at $z=L_z$. For this study, the value $L_z = 1mm$ has been chosen, as this is approximately the capsule radius. $f > 1$ is also possible, with greater field strength at the capsule center, and is therefore described here as an 'anti-mirror'.

The mirror parameter $f$ has been varied in 2D Gorgon simulations, as well as the field amplitude $B_{M0z}$. Figure \ref{fig:Tamp_B} shows the burn averaged ion temperatures against applied magnetic field strength. Now that the magnetic field is not uniform, it is important to consider how best to quantify the applied field strength. Here the volume averaged initial magnetic field magnitude in the fuel is chosen. This definition elucidates the key physics at play, as will become apparent. A separate useful definition would be from an engineering standpoint; for example: 'what current would it take to generate such a field?' The engineering challenges in making the proposed magnetic field topologies are left for a future study. 

Figure \ref{fig:Tamp_B} shows that the mirror topology also has diminishing gains as the magnetic field strength is high enough. However, the maximum temperature amplification is higher for a more severe mirror shape (lower $f$). 

This behavior is simply explained by more of the heat-flow from the hot-spot being restricted, as the field lines more closely follow the surface. Figure \ref{fig:Tamp_wt} plots the temperature enhancement for $f=0.2$ against burn-averaged electron magnetization. For this specific mirror ratio, the following definition for $\kappa_{\text{eff}}$ has been found to match the data reasonably well:

\begin{equation}
	\kappa_{\text{eff,mirror}} = 0.2 + 0.8\frac{\kappa_{\perp}}{\kappa_{\parallel}}
\end{equation}
i.e. the heat-flow can be restricted down to 20\% rather than 33\% in the axial case.

\begin{figure}
	\centering
	\centering
	\includegraphics[scale=0.6]{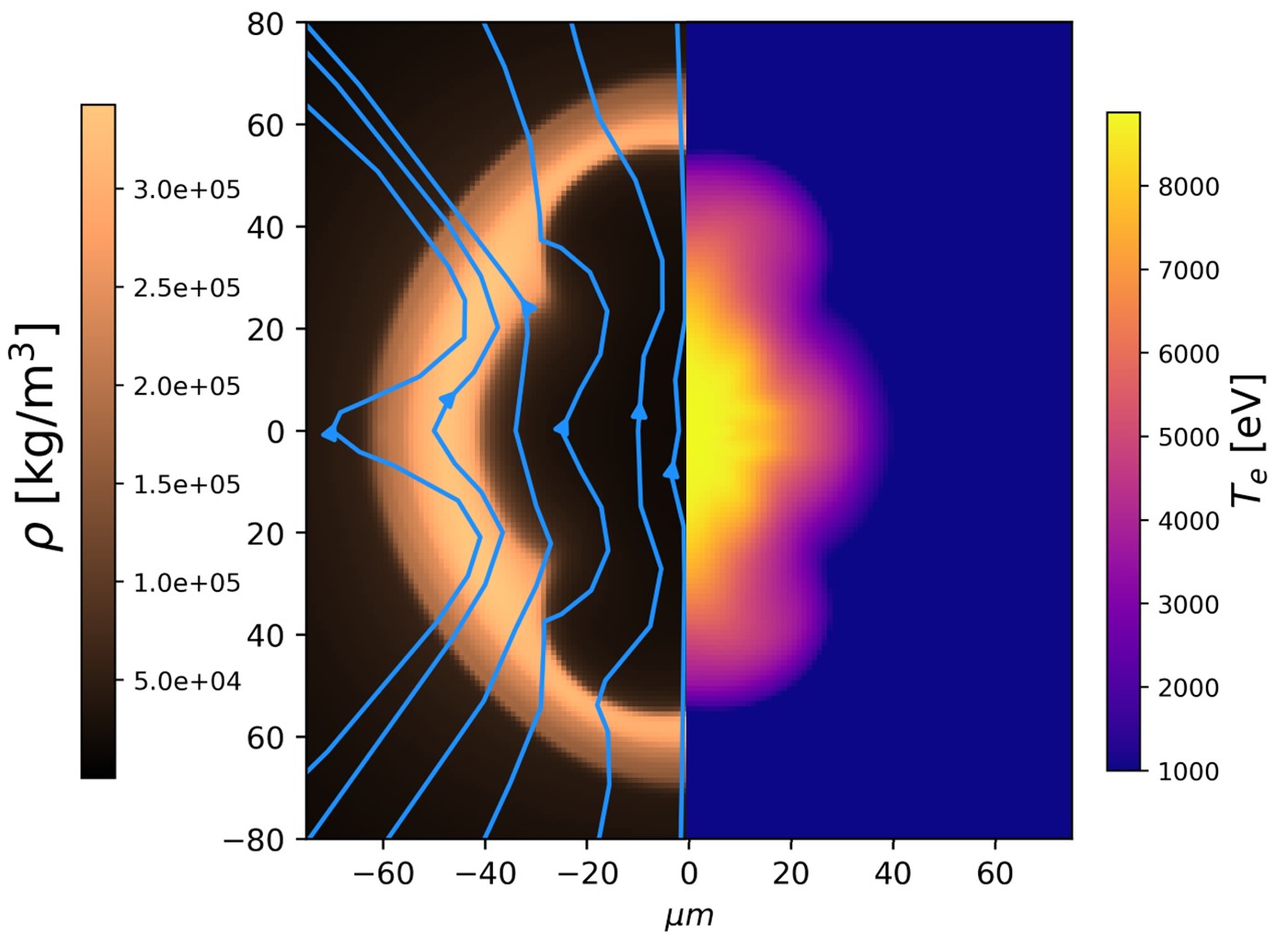}\caption{\label{fig:stag_mirror} Density, electron temperature and magnetic field streamlines at peak neutron production (t=8.5ns) for a mirror field with $f=0.2$ and $B_{M0z} = 30T$}	 
\end{figure}

Figure \ref{fig:stag_mirror} shows density, electron temperature and magnetic streamlines at stagnation (t=8.5ns) for a simulation with $B_{M0z}=30T$ and $f=0.2$, which corresponds to an initial volume averaged magnetic field strength in the fuel of 21.3T. While the magnetic field strength is lower than the 30T axial field of figure \ref{fig:stag_ax}, the peak electron temperature is around 1keV higher. 

The mirror simulations also show the hot-spot elongation that was seen for the axial field case. In addition, there is a strong P4 legendre mode that emerges. This shape asymmetry occurs due to there being 2 different regions of the capsule surface. There is a region near the waist where field lines follow the hot-spot surface. Little heat escapes the capsule in this region when strongly magnetized. Instead, it is transported along the field lines to the pole, where it readily leaves the capsule. It is possible that closer attention to the form of the mirror field would be beneficial.

\begin{figure}
	\centering
	\centering
	\includegraphics[scale=0.6]{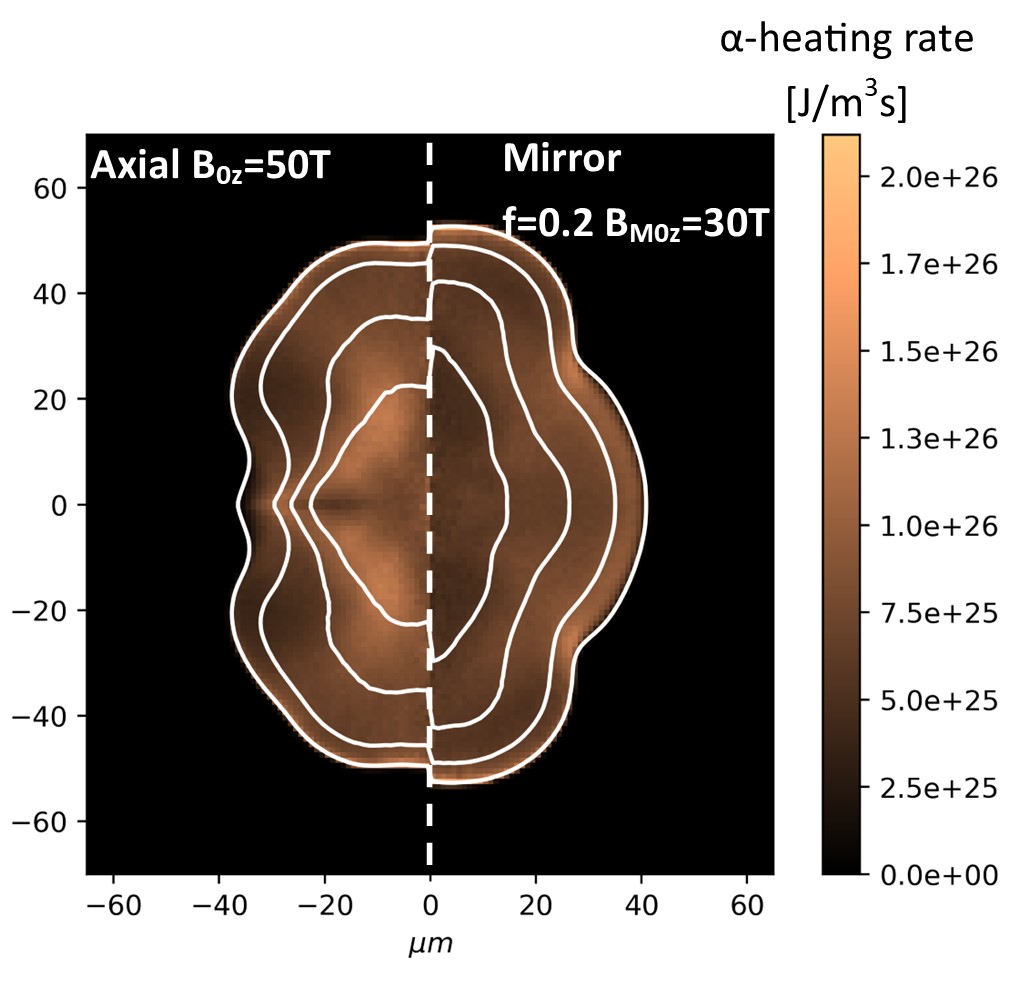}\caption{\label{fig:alpha}  $\alpha-$heating profile in the hot-spot for an axial field with $B_{0z} = 50T$ (left) and mirror field with f=0.2 $B_{M0z} = 30T$ (right). White lines are ion temperature contours of 1keV, 3keV, 5keV and 7keV.}	 
\end{figure}

So far all analysis has been without any $\alpha$-heating included. For large magnetic field strengths, the $\alpha$-heating distributions for axial and mirror fields are not noticeably different. However, as a mirror field can be used to get the same hot-spot temperature enhancement at a lower magnetic field strength (through thermal conduction suppression), a mirror field can be used to confine the $\alpha$-particles less. Figure \ref{fig:alpha} compares the $\alpha$-heating distribution at bang-time for an axial field of 50T and a mirror field using $f=0.2$ and $B_{M0z}=30T$; these had comparable temperature enhancements from thermal conduction. It can be seen that the axial case traps the $\alpha$-particles more severely to heat the core of the hot-spot, while the mirror case allows more transport to the waist for heating. Less trapping of $\alpha$-particles at the center may be desired if the magnetic field is deemed to be holding back burn wave propagation \cite{oneill2024burnpropagationmagnetizedhighyield}.  

No perturbations or mix have been included in these calculations. However, it is expected that the mirror field will enhance the impact of magnetic fields on mix, as more of the hot-spot surface normal is orthogonal to the magnetic field \cite{walshMagnetizedAblativeRayleighTaylor2022}.

A mirror field ($f<1$) can be generated by using 2 coils on the same axis to produce the magnetic field. As the coils are moved away from each other along the axis, the central field strength decreases.   

\section{Cusp Magnetic Field \label{sec:cusp}}

At the OMEGA Laser Facility 2 coils are routinely used to boost the axial field strength, with currents run through the wires in the same direction to enhance the field strength \cite{fiksel2015}. If the polarity of one coil is reversed, however, a cusp field is generated. The following magnetic field profile has been used in the Gorgon simulations:

\begin{equation}
	B_r = - \frac{r_{cyl} B_{Cu0}}{2L_z} 
\end{equation}

\begin{equation}
	B_z =  \frac{z B_{Cu0}}{L_z} 
\end{equation}
Where $L_z=1mm$ is used. This initial field results in the topology shown in figure \ref{fig:B0}. The magnetic field strength goes to zero at the capsule center. 

The cusp field is found to result in the worst performance of the topologies considered. Figure \ref{fig:Tamp_B} shows the temperature as a function of volume-averaged initial magnetic field strength; again the temperature plateaus, but at a lower peak temperature than the axial field case. 

Figure \ref{fig:Tamp_wt} then plots this against electron magnetization. The lower performance can be attributed to the fact that the cusp field has poor confinement of heat at the capsule waist, unlike the other topologies. 

\begin{figure}
	\centering
	\centering
	\includegraphics[scale=0.6]{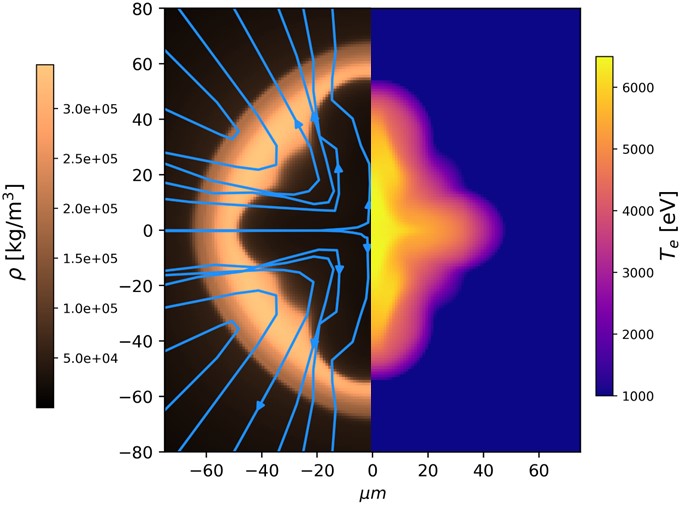}\caption{\label{fig:stag_cusp} Density, electron temperature and magnetic field streamlines at neutron bang-time (t=8.5ns) for a cusp field with $B_{Cu0} = 30T$. }	 
\end{figure}

The stagnated hot-spot shape is also a cause for concern, with density and electron temperature plotted in figure \ref{fig:stag_cusp} for $B_{Cu0} = 30T$. A strong P4 Legendre mode emerges, as well as the P2 observed in axial field configurations. In fact, a series of simulations were run to tune the shape of the implosion with P2 and P4 radiation drive asymmetries, but this showed the emergence of a P8 asymmetry in the hot-spot.

\section{Closed Magnetic Field \label{sec:closed}}

Closed magnetic field lines have been deliberated for many years as a concept to increase the impact of magnetization \cite{hohenbergerInertialConfinementFusion2012}. To the authors' knowledge, the following are the first MHD simulations of spherical ICF implosions with closed field lines in a realistic 2D geometry.

The following linear ramp of magnetic field strength is used in the simulations:

\begin{equation}
	B_{\theta} = B_{Cl0} \frac{r_{cyl}}{L_r} \label{eq:B0_closed}
\end{equation}
Where $B_{Cl0}$ is a constant that is varied in the simulations. $L_r = 1mm$ is used here. This form is used to satisfy the constraint that $B_{\theta} = 0$ at $r=0$. An alternative form was also investigated, using the magnetic field strength as if a wire was running through the capsule; this form has a linear ramp as in equation \ref{eq:B0_closed} up to the wire radius, then decaying as $1/r$. The wire solution results were broadly similar to the linear ramp simulations used here. 

Figure \ref{fig:Tamp_B} shows the burn averaged temperature against initial volume averaged magnetic field strength. Again, the enhancement of temperature with magnetization is seen to plateau at high applied field strengths, albeit at a much higher temperature than any of the other magnetic field topologies investigated in this paper. 

Extending the theory of axial thermal conduction suppression to this case, the following effective thermal conductivity can be defined:

\begin{equation}
	\kappa_{\text{eff,closed}} = \frac{\kappa_{\perp}}{\kappa_{\parallel}}
\end{equation}
Which goes to zero at large magnetic field strengths. Clearly the scaling of $T\sim\kappa_{\text{eff}}^{-2/7}$ cannot be maintained in this regime. 

There are several factors that break the theoretical scaling. First, the scaling is formulated by assuming the initial hot-spot density is zero and that the final density is all from plasma ablated into the hot-spot by thermal conduction. As the temperature is predicted to go to infinity, the stagnated hot-spot density goes to zero. In reality, the stagnated hot-spot density cannot be lower than the initial density. Also, the early-time ablation occurs before the hot-spot has become strongly magnetized. 

Secondly, the theory only takes into account thermal conduction as an energy transport process. In reality the radiation, ion conduction and $\alpha$-particle transport also drives ablation of cold fuel into the hot-spot. 

\begin{figure}
	\centering
	\centering
	\includegraphics[scale=0.6]{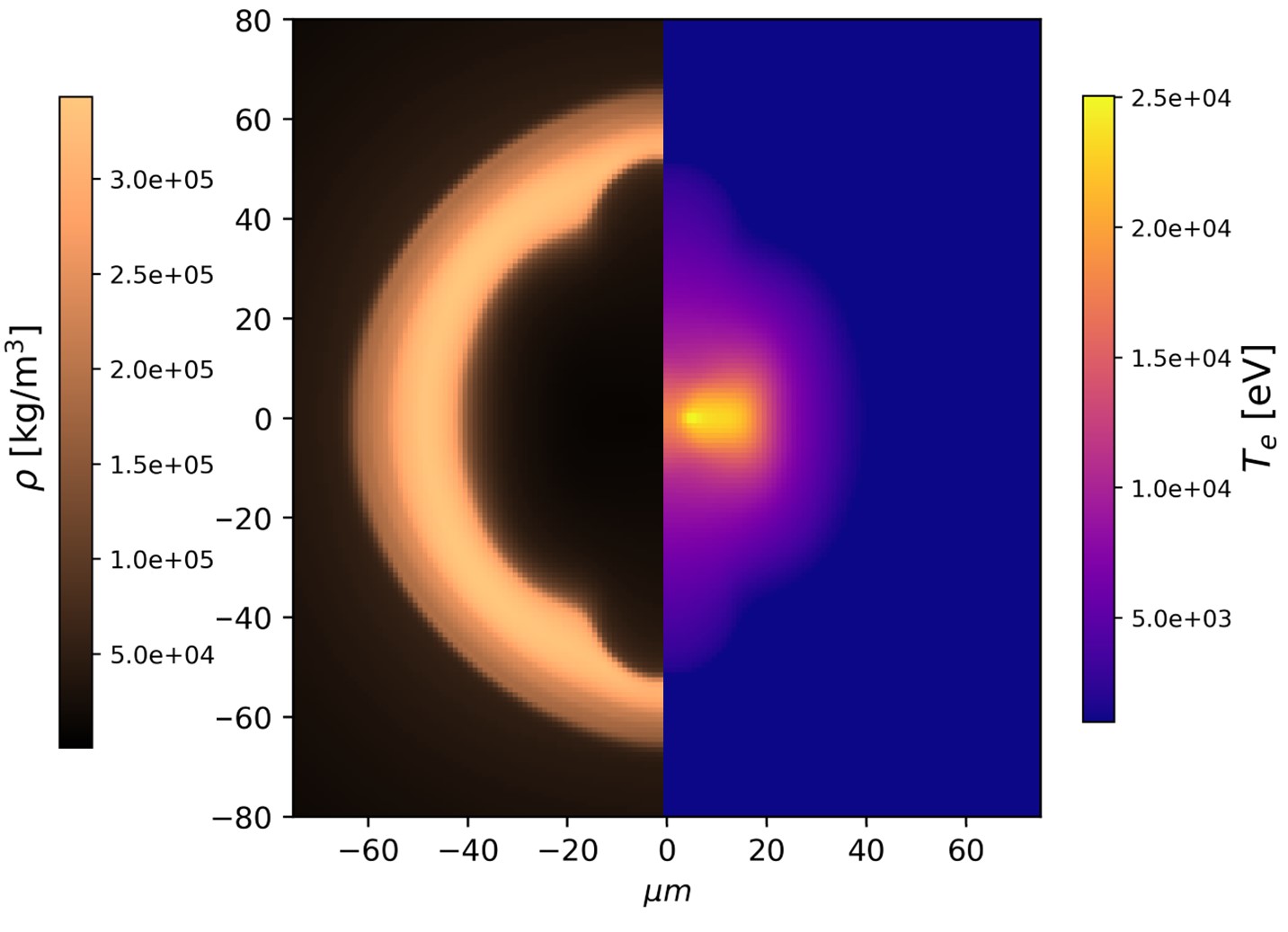}\caption{\label{fig:stag_shear1} Stagnated density and electron temperature for a case with an applied closed field line $B_{Cl0} = 30T$.}	 
\end{figure}

Finally, while closed fields can conceptually inhibit all thermal conduction from a spherical hot-spot, the constraint that $B_\theta = 0$ at $r=0$ means that heat is still lost towards the poles. In the simulations run here, this results in a significant shape asymmetry at the pole, as seen in figure \ref{fig:stag_shear1}.

Figure \ref{fig:Tamp_wt} plots the temperature amplification scaling and 2D simulation results against electron magnetization. While the scaling is unrealistic and quickly leaves the bounds of the plot, it provides an understanding of how the closed magnetic field results are able to get much larger temperature enhancements than the other topologies in this paper. 

\begin{figure}
	\centering
	\centering
	\includegraphics[scale=0.6]{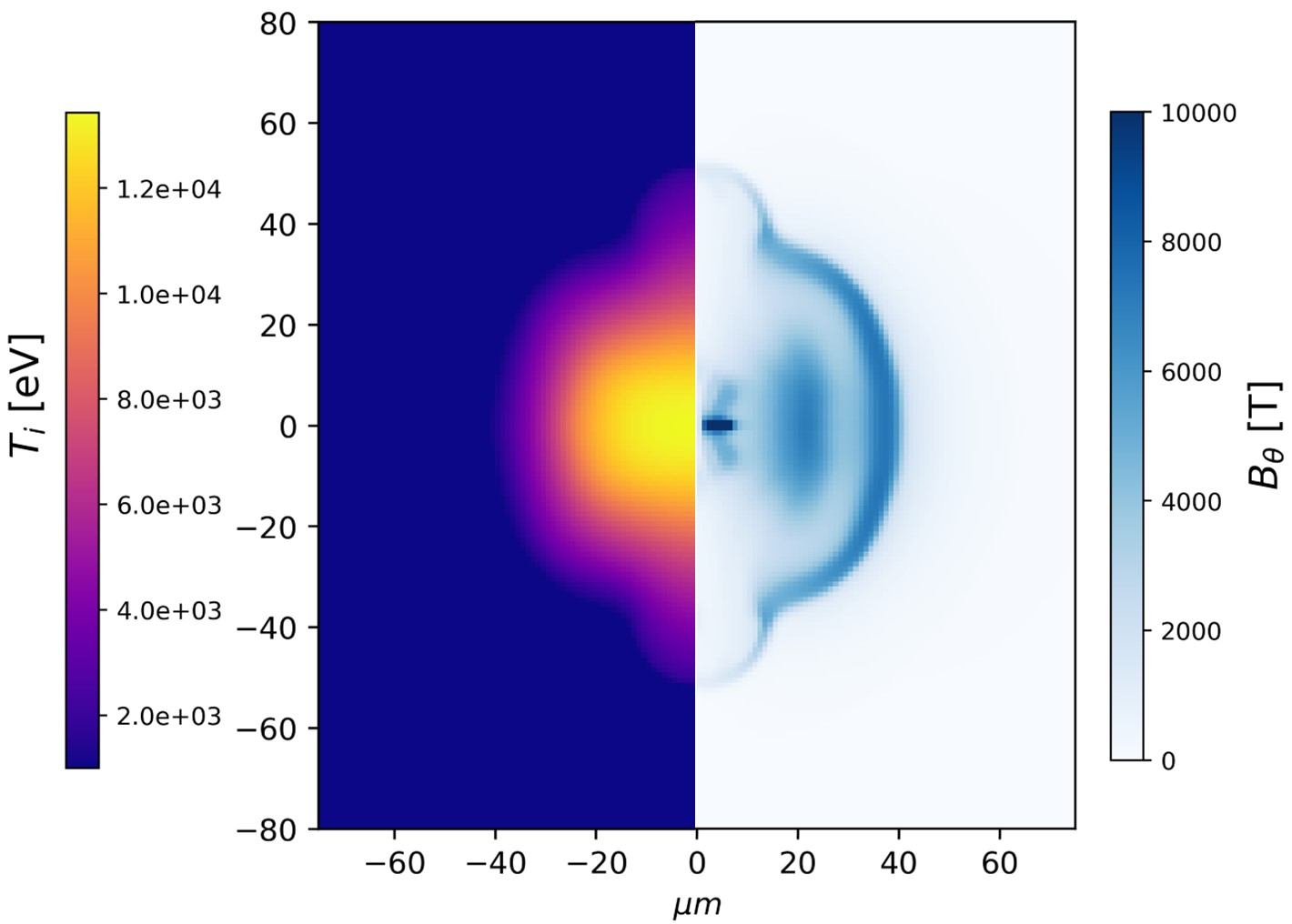}\caption{\label{fig:stag_shear2}Ion temperature and azimuthal magnetic field strength at peak stagnation (t=8.5ns) for a capsule with an imposed $B_{Cl0} = 30T$ closed field profile.}	 
\end{figure}

In fact, the burn averaged ion temperature does not capture the full impact of closed magnetic field lines. Figure \ref{fig:stag_shear1} shows the electron temperature in the hot-spot reaching 25keV. This would be an unprecedented value, even if there was $\alpha$-heating occurring in an igniting capsule. The ion temperature (plotted alongside the magnetic field strength in figure \ref{fig:stag_shear2}) is a factor of 2 lower than the electron temperature, due to the electron-ion equilibration being too slow. More than anything, these results demonstrate the critical importance of electron thermal conduction to the functioning of ICF implosions. 

\begin{figure}
	\centering
	\centering
	\includegraphics[scale=0.5]{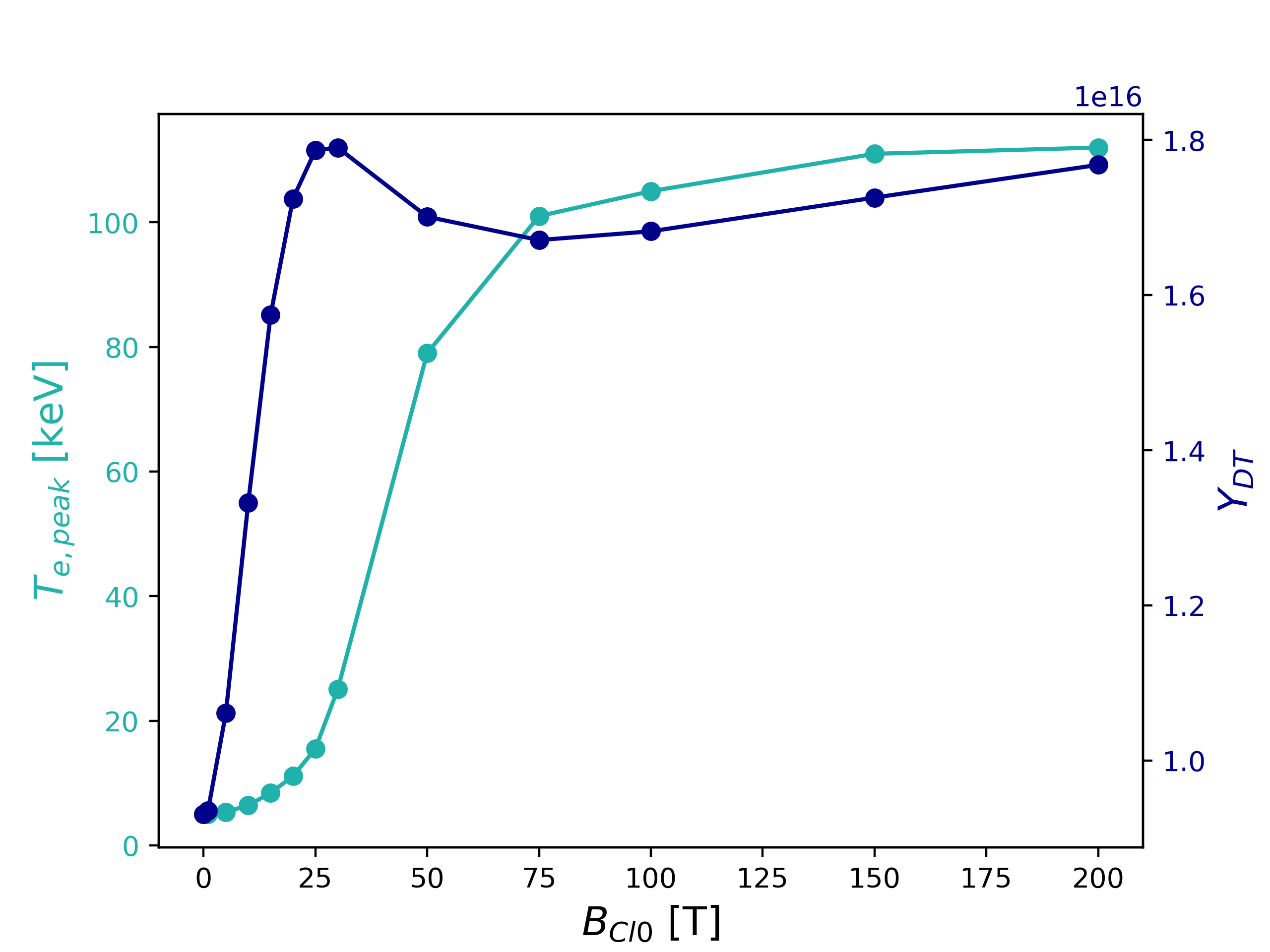}\caption{\label{fig:Te_Y_closed} Peak electron temperature and DT neutron yield versus applied closed magnetic field strength. All cases are without $\alpha$-heating and just represent the changes from magnetization of thermal conduction.}	 
\end{figure}

Figure \ref{fig:Te_Y_closed} shows the peak electron temperature and DT yield (without $\alpha$-heating) against the applied closed field strength. Electron temperatures in excess of 100keV are simulated. This result neglects non-local transport effects that would be prevalent in this regime.

The greatest difficulty for closed field lines is finding a practical method of generating the magnetic field. A current-carrying wire run down the capsule center could generate this topology \cite{hohenbergerInertialConfinementFusion2012}; however, the wire material mixing into the capsule fuel would surely result in a substantial performance degradation through radiative losses (plus pre-heat of the fuel). Another suggestion has been an '$\Omega$-coil' \cite{hohenbergerInertialConfinementFusion2012}, although this requires magnetic reconnection upon capsule compression to result in closed field lines; this is not a process that has been substantiated for capsule implosions.

As part of this study, a radiofrequency field driven down the hohlraum axis has been investigated. Simplified calculations suggest that magnetic fields increasing with radius up to $B_{Cl0} =10T$ within the capsule may be possible by driving resonant modes. Further investigation of this technique is left for future work. 

It is worth noting that all results here are for spherical targets. Cylindrical implosions with axial magnetic fields have been executed on the OMEGA Laser Facility \cite{hansenNeutronYieldEnhancement2020,bailly-grandvaux2024}. While these are not closed field lines, the field is everywhere along the hot-spot surface and therefore $\kappa_{\text{eff}}=0$ when the cylinder length is much greater than its radius.

\section{Twisted Magnetic Field \label{sec:twist}}

Another topology to consider is that of a twisted field, which is conceptually a combination of an axial field and a closed field. This is appealing, as it may be possible to impose an axial field and have it twisted during the experiment to exhibit the benefits of closed fields discussed in section \ref{sec:closed}. 

Field twisting was initially found to be inherent to magnetized implosions due to the cross-gradient-Nernst effect\cite{walshExtendedMagnetohydrodynamicEffects2018}. However, developments in the transport coefficients \cite{sadlerSymmetricSetTransport2021,daviesTransportCoefficientsMagneticfield2021} revealed these results to be erroneous. Updated transport coefficients then found a significantly reduced effect \cite{2021}.

Biermann Battery inherently generates closed fields within the capsule \cite{walshNonlinearAblativeRayleighTaylor2023}. However, the fields are largest in strength when the capsule is perturbed with large amplitude, short wavelength perturbations \cite{walshBiermannBatteryMagnetic2021}, which are not desirable for achieving high yields.

Simulations of vortical hot-spot flows during hot-spot stagnation have been shown to induce at least mild magnetic field twisting \cite{walshPerturbationModificationsPremagnetisation2019}. However, these required substantial low-mode capsule asymmetries, which are not appealing for achieving high fusion yields. 

Another possibility would be to induce twisting during the capsule drive phase through non-radial plasma velocities.  However, a back-of-the-envelope calculation shows this to be flawed. For an initial axial field, an azimuthal velocity gives the following development of the twisted field component (in cylindrical coordinates):

\begin{equation}
	\frac{\partial B_\theta}{\partial t} = B_z  \frac{\partial v_\theta}{\partial z} 
\end{equation}
Large non-radial velocities at the capsule waist, while keeping the pole only imploding radially, would result in twisting. An approximation for the relative strength of the closed to axial field component gives:
\begin{equation}
	\frac{B_\theta}{B_z} \sim \frac{t v_\theta}{r}
\end{equation}
i.e. the non-radial plasma velocity would have to be of the same order as the implosion velocity to induce substantial twisting. This is infeasible, first and foremost due to the large kinetic energy that would not contribute to stagnation.

\section{Conclusion}

The imposition of 4 magnetic field topologies have been compared. The highest performance comes from applying closed magnetic field loops within the capsule fuel. Simulations here show up to 2x ion temperature increase from magnetization. However, this under-represents the impact of closed fields; electron temperatures in excess of 100keV have been simulated, without the inclusion of $\alpha$-heating. This points to the possibility of radically redesigning ICF implosions to make the most of closed field magnetization effects. 

Mirror fields have also been found to exceed the performance benefits of an axial field, due to the field lines following the hot-spot surface more closely. Temperature enhancements in excess of 60\% have been simulated. These results correspond to an effective thermal conductivity of 0.2. 

Cusp fields, on the other hand, have not been found to have any benefits over an axial magnetic field. The shape perturbation is more severe and there are lower levels of heat-flow suppression. 

All of the results were taken in terms of temperature enhancement, due to the fact that the shape asymmetries inherent to magnetization have not been tuned out; the temperature is less sensitive than the yield to these asymmetries \cite{walshMagnetizedICFImplosions2022}. The temperature enhancements simulated here can be applied to the yield scaling in \cite{walshMagnetizedICFImplosions2022} to estimate the yield enhancement if the implosions were symmetric. The result is then dependent on the unmagnetized hot-spot temperature, with cooler hot-spots benefitting the most from magnetization. For a 2keV initial hot-spot, an axial field increases the yield by 1.9$\times$. A mirror field that gives 60\% temperature amplification corresponds to a 2.5$\times$ yield enhancement by magnetization. A closed field that gives 100\% temperature enhancement would then increase the yield by 3.5$\times$.


\section*{Acknowledgements}
This work was performed under the auspices of the U.S. Department of Energy by Lawrence Livermore National Laboratory under Contract DE-AC52-07NA27344. 

Work supported by LLNL LDRD project 23-ERD-025.

This document was prepared as an account of work sponsored by an agency of the United States government. Neither the United States government nor Lawrence Livermore National Security, LLC, nor any of their employees makes any warranty, expressed or implied, or assumes any legal liability or responsibility for the accuracy, completeness, or usefulness of any information, apparatus, product, or process disclosed, or represents that its use would not infringe privately owned rights. Reference herein to any specific commercial product, process, or service by trade name, trademark, manufacturer, or otherwise does not necessarily constitute or imply its endorsement, recommendation, or favoring by the United States government or Lawrence Livermore National Security, LLC. The views and opinions of authors expressed herein do not necessarily state or reflect those of the United States government or Lawrence Livermore National Security, LLC, and shall not be used for advertising or product endorsement purposes.

The data that support the findings of this study are available from the corresponding author upon reasonable request.

	\section*{Bibliography}
	
	\bibliographystyle{unsrt}

\end{document}